\documentclass[twocolumn,prb,aps,showpacs,floatfix,superscriptaddress]{revtex4}
\usepackage{graphicx}
\usepackage{verbatim}

\begin{document}

  \newcommand{\wert}[3]{$#1=#2\,\rm{#3}$}
  \newcommand{\LSMO} {La$_{2-2x}$Sr$_{1+2x}$Mn$_2$O$_7$ }
  \newcommand{\LSMOE} {La$_{2-2x}$Sr$_{1+2x}$Mn$_2$O$_7$}
  \newcommand{\LSMOH} {LaSr$_{2}$Mn$_2$O$_7$}
  \newcommand{\LSMOL} {La$_{1.28}$Sr$_{1.72}$Mn$_2$O$_7$}
  \newcommand{\rangeS}[2]{$#1\,\rm{eV} \leq \Delta E \leq #2\,\rm{eV}$}
  \newcommand{\rangeL}[2]{$#1\,\rm{keV} \leq \Delta E \leq #2\,\rm{keV}$}

\title{\textit{d$\,$-$\,$d} Excitations in Bilayer Manganites Probed by Resonant Inelastic X-ray Scattering}

\author{F. Weber}
\email{frank.weber@kit.edu}
\altaffiliation{present address: Karlsruhe Institute of Technology, Institute of
Solid State Physics, P.O. Box 3640, D-76021 Karlsruhe, Germany\\ \\}
\affiliation{Materials Science Division, Argonne National Laboratory, Argonne, IL, 60439, USA}

\author{S. Rosenkranz}

\affiliation{Materials Science Division, Argonne National Laboratory, Argonne, IL, 60439, USA}

\author{J.-P. Castellan}

\affiliation{Materials Science Division, Argonne National Laboratory, Argonne, IL, 60439, USA}

\author{R. Osborn}

\affiliation{Materials Science Division, Argonne National Laboratory, Argonne, IL, 60439, USA}

\author{J. F. Mitchell}

\affiliation{Materials Science Division, Argonne National Laboratory, Argonne, IL, 60439, USA}

\author{H. Zheng}

\affiliation{Materials Science Division, Argonne National Laboratory, Argonne, IL, 60439, USA}

\author{D. Casa}

\affiliation{Advanced Photon Source, Argonne National Laboratory, Argonne, IL, 60439, USA}

\author{J. H. Kim}

\affiliation{Advanced Photon Source, Argonne National Laboratory, Argonne, IL, 60439, USA}

\author{T. Gog}

\affiliation{Advanced Photon Source, Argonne National Laboratory, Argonne, IL, 60439, USA}

\begin{abstract}
  We report a high resolution resonant inelastic x-ray scattering investigation of the bilayer manganites La$_{2-2x}$Sr$_{1+2x}$Mn$_2$O$_7$ with $x = 0.36$ and $0.5$. The momentum dependence along the crystallographic $(110)$ direction for energy losses $1\,\rm{eV} \leq \Delta E \leq 15\,\rm{eV}$ has been measured in detail with the data analysis focusing on the energy loss region $1\,\rm{eV} \leq \Delta E \leq 5\,\rm{eV}$, which includes a strong peak located at $\Delta E \approx 2\,\rm{eV}$.  We observe a clear dispersion of up to $0.5\,\rm{eV}$ in the measured $\textbf{q}$ range, which is direct evidence of the non-local character of this excitation. Further, we found that the intensity in this low energy region strongly depends on both the reduced wave vector $\textbf{q} = (h, h, 0)$, $h = 0.1 - 0.5$, and temperature, i.e. different ordered phases. Results can be explained via an intersite $d-d$ charge transfer excitation, proposed for pseudo-cubic manganites, where the hopping rate is strongly increased (decreased) by ferromagnetic (antiferromagnetic) alignment of neighboring in-plane Mn ion core spins.
\end{abstract}


  \maketitle

  \vskip2pc

\section{Introduction}
\label{intro}

The complex phase diagrams of many transition metal oxides highlight the strong interplay and competition between lattice, spin and charge degrees of freedom\cite{Schiffer95,Mitchell01,Li07,Bonn06}. The phase diagram of the bilayer manganite \LSMO is among the most complex ones and still subject to refinement\cite{Li07,Li06}. In particular, the CE-type antiferromagnetic charge and orbital ordered ground state of the half-doped bilayer manganite \LSMOH, predicted by Goodenough\cite{Goodenough55} for half doped manganites in 1955, was only recently observed\cite{Li07}. In this state, the $e_g$ charge is localized on alternate Mn sites producing to first order distinct Mn$^{3+}$ and Mn$^{4+}$ sites\cite{Radaelli97,Argyriou00}. The Jahn-Teller effect and orbital ordering of the $e_g$ charge induces a characteristic distortion with a superlattice peak at $\textbf{q} = (\frac{1}{4}, \frac{1}{4}, 0)$. For lower-doped bilayer manganites $(0.3 \leq x \leq 0.4)$, the frustration of the charge and orbital order results in a ferromagnetic metallic ground state with a transition to a paramagnetic insulator at elevated temperatures. Above $T_C$, the appearance of diffuse elastic and quasi-elastic diffraction peaks in the vicinity of $\textbf{q}_{CE}$ (Ref.~\onlinecite{Argyriou02}) shows that $15$-\AA$\,$ sized charge and orbitally ordered clusters coexist with charge-delocalized clusters\cite{Sun07}. At $T_C$, the onset of long-range ferromagnetic order quickly melts the short-range charge and orbital correlations and leads to colossal magnetoresistance\cite{Vasiliu99}. The low temperature state of  La$_{1.2}$Sr$_{1.8}$Mn$_2$O$_7$ ($x = 0.4$) is characterized by strong electron-phonon coupling observed by ARPES\cite{Manella05} and inelastic neutron scattering, which identified strongly renormalized Jahn-Teller-type optical phonon modes at $\textbf{q}_{CE}$ (Ref. \onlinecite{Weber09}). Thus, the ferromagnetic metallic state of this bilayer manganite offers the defining characteristics of a polaronic metal, yet another exotic and poorly understood state in the rich phase diagram of the bilayer manganites.

Typical theoretical approaches for the manganites neglect the oxygen degrees of freedom and try to parameterize the behavior of the Mn ion in terms such as the intersite hopping amplitude of the $e_g$ electrons and on-site Coulomb repulsion, e.g. HundÕs coupling. Measurements of electronic excitations in the energy range of these parameters, typically several eV, are therefore particularly desirable and can provide critical information for new theoretical approaches.

Resonant inelastic x-ray scattering (RIXS) at transition metal $K$-edges is a bulk-sensitive, momentum-resolved probe of electronic excitations in the order of eV and thereby has become a versatile tool in the study of strongly correlated solids. It is a photon-in-photon-out process, in which the incoming photon creates a short-lived exciton comprising a $1s$ core hole and a $4p$ photoelectron. The exciton strongly interacts with the valence electrons. The outcoming photon, created in the recombination of the $1s$ core hole and the $4p$ photoelectron, differs from the incoming one by the energy and momentum transferred to the valence electrons. The cross section of this exciton-valence electron interaction depends strongly on the incident energy.

RIXS, in particular, enables the study of so-called $d-d$ excitations in transition metals between the highest occupied and lowest unoccupied electronic state. In manganites, this is the excitation from the lower to the upper of the two $e_g$ orbitals, the properties of which are supposed to govern many of the physical properties. We note that, although in principle both intrasite and intersite $d-d$ excitations are possible, so far RIXS results have always been interpreted in the picture of an intersite $d-d$ excitation. However, direct experimental evidence of the non-local character, i.e. a dispersion, has not always been observed. In particular, results obtained for manganites were explained in terms of temperature dependent intensity variations. Intersite $d-d$ excitations, particularly in charge and orbital ordered manganites, may give important insights into the microscopic mechanisms leading to the observed exotic properties.

Studies of the electronic excitation spectrum in manganites have focused so far on the pseudo-cubic perovskites La$_{1-x}$(Ca/Sr)$_x$MnO$_3$. It was argued\cite{Inami03} that a peak at around $2.5\,\rm{eV}$ in orbital ordered LaMnO$_3$ corresponds to a Mn $d-d$ transition. The doping dependence was investigated in La$_{1-x}$Sr$_x$MnO$_3$ (Ref.~\onlinecite{Ishii04}) and Grenier et al.\cite{Grenier05} studied the influence of different magnetic ground states on RIXS intensities for energy losses from $1\,\rm{eV}$ to $5\,\rm{eV}$ in four different pseudo-cubic manganites. Based on model calculations they concluded that increased (decreased) intensities at these energy losses on entering a ferromagnetic (antiferromagnetic) state are explained in terms of an intersite $d-d$ excitation. On the same basis, they ruled out any significant spectral weight from intrasite $d-d$ excitations in their RIXS spectra. We want to point out that the non-local character was purely derived from the intensity variations and model calculations. No wave vector dependence of the excitation energy, which is a direct experimental evidence of the non-local character of an excitation, was reported.

 \begin{table}
        \centering
            \begin{tabular}{c|c|c}
            \label{table1}
            \textbf{Compound} &  \textbf{T (K)} &  \textbf{Phase} \\\hline
            \LSMOL ($x\,=\,0.36$) & 20   & FM - Metal\\
                                                    & 150 & PM short-range COO\\\hline
	  \LSMOH ($x\,=\,0.5$)   & 75   & AFM(CE) COO\\
	                 (sample CE)   & 175   & PM COO\\
                                                    & 250 & PM\\\hline
	  \LSMOH ($x\,=\,0.5$)   & 20   & AFM(A)\\
	                       (sample A)& 160   & AFM(CE) COO\\
	                                           & 250   & PM	
            \end{tabular}
        \caption{List of samples and the measured temperatures plus the corresponding phases. Phases are insulating if not stated otherwise. The following abbreviations are used: FM/PM/AFM for ferro-/para-/antiferromagnetic; COO for charge and orbital ordered; AFM(CE) and AFM(A) for CE- type and A-type antiferromagnetism, respectively; Sample CE has a CE-ordered groundstate. Sample A is CE-ordered only in a limited temperature range and features an AFM(A) ground state.}

    \end{table}

Here, we present a RIXS investigation at the Mn $K$-edge of the bilayer manganite \LSMO with $x = 0.36$ and $0.5$. The lower doped sample shows colossal magnetoresistance near a ferromagnetic metal - insulator transition at \wert{T_C}{130}{K} as described above. For the half-doped composition, we utilize two samples, both of which enter the CE-type charge and orbital ordered state upon cooling. However, only in one of them, sample CE, this is the ground state. The second specimen, sample A, undergoes a transition around \wert{T}{75}{K} to an A-type antiferromagnetic phase. In a recent publication\cite{Li07} it was shown that extremely small variations in the doping, which cannot be quantified by measurement of the doping concentration, are at the origin of these different ground states at nominally equal doping levels. The authors tentatively assigned an exact doping level of $x = 0.5$ to samples with the CE-type ordered ground state\cite{Li07}. The other, so-called reentrant samples were assigned to a slightly lower or higher doping value without having a quantitative number for the difference. Above \wert{T}{225}{K} both samples with nominally $x = 0.5$ are paramagnetic insulators. The sample with the CE-ordered ground state features also a CE-type charge and orbital ordered paramagnetic phase between \wert{T}{150}{K} and $225\,\rm{K}$. A list with all measured temperatures and corresponding phases is given in table ~\ref{table1}.

The presentation of our study proceeds as follows. In section~\ref{experimental} we give details about the sample parameters, used experimental configurations and data analysis. Results and discussion of the RIXS measurements are given in section~\ref{results}. Here, we focus on the intensities at energy losses $\Delta E \leq 5\,\rm{eV}$ and their \textbf{q} dependence. The paper concludes with remarks on open issues of RIXS in bilayer manganites.

\section{Experimental}
 \label{experimental}

  Our samples were all high-quality single crystalline platelets with dimensions of about $(4$x$3$x$0.5)\,\rm{mm}^{3}$ and the crystallographic c-axis perpendicular to the large surface of the platelets. Throughout this publication we use the undistorted tetragonal structure ($I4/mmm$) with $a = b \approx 3.88\,$\AA$\,$ and $c \approx 19.8\,$\AA$\,$ at all temperatures. The components of wave vectors $\textbf{Q} = (H, K, L)$ and $\textbf{q} = (h, k, l)$ are given in reciprocal lattice units ($r.l.u.$) of ($\frac{2\pi}{a}$, $\frac{2\pi}{b}$, $\frac{2\pi}{c}$).

The experiments were carried out at beamline 9ID-B at the Advanced Photon Source, Argonne National Laboratory. X-rays from a standard undulator 'A' were monochromated by a main Si(111) and secondary Si(311) monochromator. The spectrometer has a vertical scattering geometry.

During our data collection, various analyzer/detector configurations were used as they became available, ranging from a basic finely diced spherical 1-m Ge(531) analyzer with an amptek point detector (as used in Ref. 15) to a full dispersive setup microstrip detector with a large-dice Si(440) 1-m spherical analyzer\cite{Casa10}.

These different setups provided increasing energy resolution and throughput. While the conclusion drawn from spectra collected do not depend on the particular setup, the figures do show a marked difference in count rates in some cases, due to the increased efficiency of the Si(440) analyzer/microstrip detector configuration.

   \begin{figure}
   \includegraphics[width=0.9\linewidth]{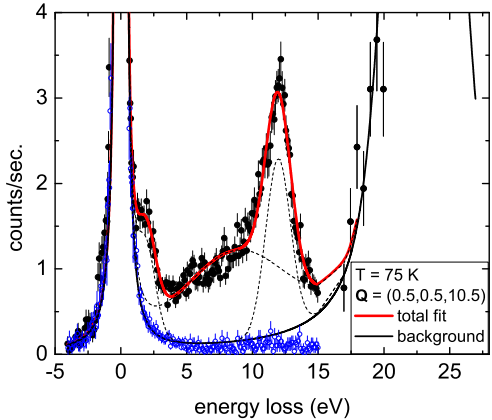}
   \caption{(color online) Inelastic scattering measured with an incident energy above, \wert{E_i}{6.557}{keV} (black dots), and below, \wert{E_i}{6.505}{keV} (blue circles), the Mn $K$-edge in CE-ordered \LSMOH ($x = 0.5$, sample CE) at \wert{T}{75}{K}. The solid red line represents the total fit of the resonant data set consisting of a fit of the background (solid black line, incorporating the elastic ($\Delta E = 0$) and $k_{\beta}$ emission lines ($\Delta E \approx 23\,\rm{eV}$)) and three Gaussians fitted to the resonantly enhanced part for energy losses $1\,\rm{eV} \leq \Delta E \leq 15\,\rm{eV}$ (dashed black lines on top of the background function).}
   \label{fig_1}
  \end{figure}

In particular, analyzer reflections used for \LSMOL /  \LSMOH (sample CE) /  \LSMOH (sample A) were Ge(531) / Ge(531) / Si(440) producing a resolution of $450\,\rm{meV}/210\,\rm{meV}/230\,\rm{meV}$ (FWHM of elastic line), the first with point detector and last two with microstrip detector. The count rates, while not directly comparable, were much improved, especially in the last setup.
The incident energy was tuned to \wert{E_i}{6.557}{keV}, above the Mn $K$-edge ($6.539\,\rm{keV}$), where the RIXS intensity was maximized.  This energy was used for all spectra shown unless otherwise stated. Resonant inelastic scans were performed at the temperatures given in table \ref{table1}.

Figure~\ref{fig_1} shows our analysis of an energy loss spectrum in the ground state of sample CE at $\textbf{Q} = (h, h, 10.5)$, $h = 0.5$. We chose to measure at these wave vectors because $\textbf{Q} = (0, 0, 10)$ is an allowed Bragg reflection and strong elastic scattering is observed throughout the adjacent Brillouin zone. Although the $(0, 0, 11)$ reflection is structurally not allowed in a perfect crystal, strong truncation rod scattering was seen in x-ray diffraction along the c* axis in La$_{1.2}$Sr$_{1.8}$Mn$_2$O$_7$ (Ref.~\onlinecite{Vasiliu99}). Scattering with a reduced c*-axis component $l = 0.5$ results in a much weaker elastic line and lowers the minimum energy loss, for which we can extract reliable RIXS intensities. The corresponding scattering process takes place between equivalent atomic positions in different unit cells and, therefore, our reduced scattering vector can be regarded as in-plane scattering. It is also generally assumed that RIXS does not obey Brillouin zone dependent structure factors. Therefore, we will present all evaluations as a function of the reduced scattering wave vector $\textbf{q} = (h, h, 0)$.

The raw data were normalized to monitor and in a second step to the integrated intensity of the $k_{\beta}$-emission line, which is located at \rangeS{17}{27}. The elastic and the $k_{\beta}$-emission lines were fitted using a Lorentzian lineshape convoluted with a Gaussian. This function provided the best fits with a consistent set of parameters. Data up to $0.6\,\rm{eV}$ in energy loss were used for fitting the elastic line. A comparison with the observed non-resonant data corroborates the use of the fit-function as the estimated background. We note that the non-resonant data were measured at a slightly lower $\textbf{Q} = (0.1, 0.1, 10.5)$. However, the footprint of the sample on the analyzer does not change significantly in the measured \textbf{Q} range and, correspondingly, we did not observe any wave vector dependence in our non-resonant scans. The resulting resonantly enhanced spectral weight for \rangeS{1}{15} were analyzed in terms of three Gaussian peaks (dashed lines): one for the Ò$2\,\rm{eV}$Ó feature, another for the peak around $\Delta E = 12\,\rm{eV}$ and a broad function describing the resonant scattering at intermediate energy transfers.

\section{Results and Discussion}
\label{results}

  \begin{figure}
   \includegraphics[width=0.9\linewidth]{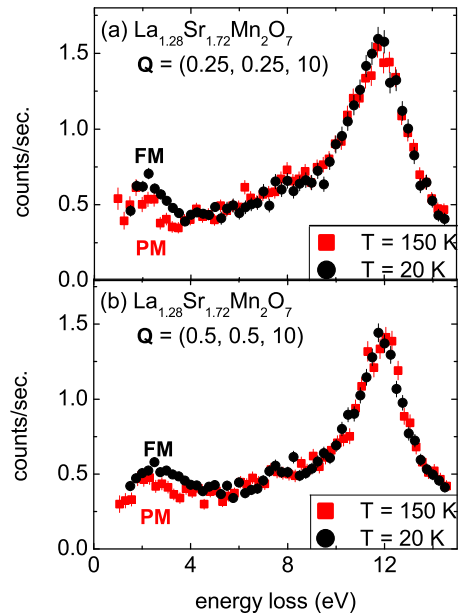}
   \caption{(color online) Background subtracted RIXS spectra in \LSMOL ($x = 0.36$). Each panel features scans above (squares) and below (dots) the FM transition temperature \wert{T_C}{130}{K} at the same wave vector along the $\Gamma Ð M$ line. No temperature dependence is observed except for the peak around $\Delta E \approx 2\,\rm{eV}$, which shows an increased spectral weight in the FM state.}
   \label{fig_2}
  \end{figure}

  \begin{figure}
   \includegraphics[width=0.9\linewidth]{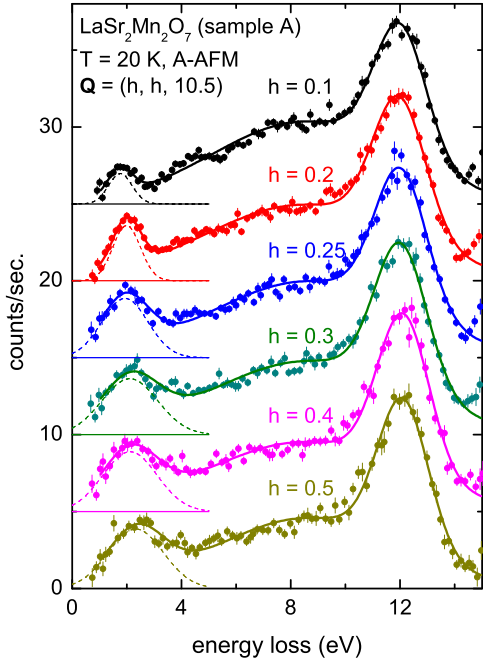}
   \caption{(color online) Background subtracted RIXS spectra in reentrant \LSMOH ($x = 0.5$) at \wert{T}{20}{K} (sample A). Thick solid lines are three-peak fits (see text and figure~\ref{fig_1}). Dashed lines show the single Gaussian fit for the low energy loss peak. The measured wave vectors are $\textbf{Q} = (h, h, 10.5)$, $h = 0.1 Ð 0.5$. An offset is included for clarity denoted by the color-coded horizontal lines.}
   \label{fig_3}
  \end{figure}

  Momentum-resolved resonant inelastic x-ray scattering at the Mn K-edge, so far, has been exclusively applied to the study of perovskite manganites of the form RE$_{1-x}$AE$_x$MnO$_3$ (RE = La, Pr, Nd; AE = Ca, Sr). Experiments\cite{Inami03,Ishii04,Grenier05} consistently reported resonantly enhanced peaks at energy losses of $2-3\,\rm{eV}$ and near $12\,\rm{eV}$. Intermediate energy loss values showed more or less structured scattering for different samples and doping, e.g. undoped LaMnO$_3$ featured a third peak at $8.5\,\rm{eV}$ (Ref.~\onlinecite{Inami03}).

We took a new approach focusing on the bilayer manganites  \LSMOE. The bilayer manganite family has the advantage of regular arranged MnO$_6$ octahedra, i.e. the Mn-O-Mn bonding angle is $180^{\circ}$. This eliminates complications from variation in the bonding angle as observed by optical spectroscopy\cite{Kim06} in pseudo-cubic manganites $R$MnO$_3$ for different $R$ ions. In particular, we choose doping values, where CE-type charge and orbital order is present, long range ($x = 0.5$) or short range ordered ($x = 0.36$).

RIXS spectra for the colossal magnetoresistive sample ($x = 0.36$) are shown in figure~\ref{fig_2}. Data above and below the Curie temperature \wert{T_C}{130}{K} are shown. In analogy to the results of perovskite manganites, we can identify two peaks at energy loss values of $2\,\rm{eV}$ and $12\,\rm{eV}$. Otherwise, the spectra are featureless. The spectra at a specific scattering wave vector are unchanged across $T_C$ except for a clear increase of the $2\,\rm{eV}$ peak intensity in the ferromagnetic state. Based on RIXS intensities at energy losses \rangeS{1}{5} measured in different magnetic phases of $30\,$\%$\, -\, 50\,$\% doped perovskite manganites, Grenier et al.\cite{Grenier05} identified the low energy loss peak in their spectra as an Mn intersite $d-d$ excitation. Their model calculation applied the stong Hund's coupling in Mn$^{3+}$ strongly favoring ferromagnetic alignment of the three $t_{2g}$ and the $e_{g}$ electrons. Thus, the intersite hopping rate of the $e_g$ electron, i.e. the $d-d$ excitationÕs spectral weight, is strongly increased by ferromagnetically aligned Mn core spins. It is strongly suppressed for antiferromagnetic order as this would require an improbable spin-flip transition. This is in qualitative agreement with our observations for \LSMOL ($x = 0.36$) indicating that our $2\,\rm{eV}$ peak represents a non-local $d-d$ excitation, too.

We also want to briefly address the nature of the second excitation present in our RIXS spectra at $\Delta E \approx 12\,\rm{eV}$. We observe no significant temperature or wave vector dependence of this peak in any of our three samples. Similar peaks were observed in RIXS measurements of pseudo-cubic manganites\cite{Inami03,Grenier05} as well as in Na$_x$CoO$_2$ (Ref.~\onlinecite{Leininger06}). Although the excitation was first assigned to O 2p $\rightarrow$ Mn 4s/4p transitions\cite{Inami03}, later reports\cite{Grenier05,Leininger06} concluded that it is a damped plasmon, i.e. a feature of $S(\textbf{Q},\omega)$.

Figure~\ref{fig_3} shows a typical set of RIXS scans taken in reentrant sample A of the two nominally half-doped bilayer manganite crystals. Equivalent data were taken at all temperatures and phases listed in table~\ref{table1} for the half-doped samples. Because we did not observe particular wave vector or temperature dependencies for the $12\,\rm{eV}$ peak, we focus our attention on the low energy loss region \rangeS{1}{5}. Here, the data taken in the A-type antiferromagnetic phase of sample A presented in figure~\ref{fig_3} show a strong increase of spectral weight for the peak near $2\,\rm{eV}$. At a closer look, we also see a clear dispersion of about $0.5\,\rm{eV}$ over the measured wave vector range. We will first discuss the intensities in the low energy loss region and then turn towards the details of the dispersion.

  \begin{figure}
   \includegraphics[width=0.9\linewidth]{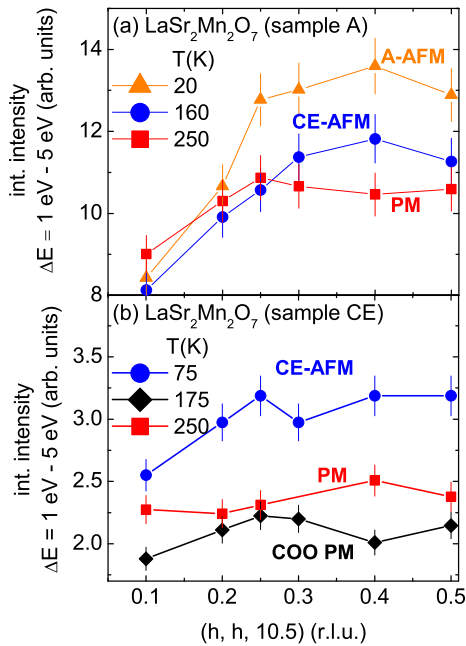}
   \caption{(color online) $\textbf{q}$ and temperature/magnetic phase dependence of the integrated intensities at energy losses \rangeS{1}{5} in (a) reentrant \LSMOH (sample A) and (b) non-reentrant \LSMOH (sample CE). Lines are guides to the eye. Error bars represent a $5\,$\% uncertainty in the background subtraction.}
   \label{fig_4}
  \end{figure}

In order to compare our low energy loss intensities with the work of Grenier et al.\cite{Grenier05}, we plot the integrated intensity of the RIXS spectra for \rangeS{1}{5} as a function of wave vector for all investigated phases for the half-doped samples in figure~\ref{fig_4}. We note that the spectral weight of the Ò$2\,\rm{eV}$Ó peak gives qualitatively the same results. Neglecting the $\textbf{q}$ dependence and focusing on $h \geq 0.25$, we see the highest integrated intensities in the antiferromagnetic phases of both samples. This seems to be in direct contradiction to the results of Grenier et al.\cite{Grenier05} and also to our own measurement in \LSMOL ($x = 0.36$), where the intensities were highest in the ferromagnetic phase.

Here, we want to remind the reader of the anisotropic nature of the bilayer manganite system expressed for example in in- and out-of-plane electrical conductivity measurements\cite{Li07,Huang08,Okimoto00}, where the conductivity $\sigma$ in bilayer manganites is closely related to the double-exchange mechanism and proportional to the hopping rate $t \propto {t_0\, cos(\frac{\eta}{2})}^{2}$ with $\eta$ being the angle between neighboring spins. Thus, conductivity measurements indicate a much bigger hopping rate in the basal plane. Further, optical spectroscopy has shown\cite{Okimoto00} that the occupied Mn$^{3+}$ $e_g$ orbital has in-plane, i.e. $d_{x^2-y^2}$ character. Therefore, we argue that in our experiments the in-plane magnetic order is much more important for the hopping rate, i.e. the spectral weight of the intersite $d-d$ excitation, than out-of-plane ordering.

The A-type antiferromagnetic phase in sample A has ferromagnetic ab-planes coupled antiferromagnetically along the c-axis. Thus, the increased intensities are explained within the model of Grenier et al.\cite{Grenier05} taking into acount only the inplane order.  The in-plane arrangement in the CE-type antiferromagnetic phase consists of ferromagnetic stripes, which are coupled antiferromagnetically along the perpendicular direction. Therefore, we expect an intermediate spectral weight of the excitation compared to the A-type antiferromagnetic and paramagnetic phases. This behavior is observed in sample A, which exhibits A-type order at low temperatures and CE-type antiferromagnetism at intermediate temperatures.

However, comparing the results for both half-doped samples we find that the gain in intensity on entering the CE-ordered phase in sample CE is bigger than on entering the A-type antiferromagnetic phase in sample A compared to the respective high-temperature paramagnetic phases. This fact is probably related to the much stronger phase competition in sample A and therefore not in contradiction to the above outlined reasoning of higher spectral weights in ferromagnetic phases.

Turning to the $\textbf{q}$ dependence of the results presented in figure~\ref{fig_4}, we see that the differences in intensities for different magnetic phases in sample A are only present for $h \geq 0.25$. If we assume the above described model captures the essential physics in terms of the spectral weight of the $d-d$ excitation, our results for $h \leq 0.2$ have to be interpreted as a loss of the influence of magnetic ordering for these wave vector values. As the scattering in our experiments has a large absolute component along c*, we cannot determine the correlation length from these results. Also, due to small sample size neutron diffraction\cite{Li07} was not able to properly determine the correlation length of the magnetic order. We see a qualitative difference in sample CE, where the maximum spectral weight for each wave vector appears for the same phase, i.e. CE-type antiferromagnetism, although the difference to the paramagnetic phases is reduced at $h = 0.1$. The spectral weight of the $d-d$ excitation in sample CE seems to be, though on a smaller scale, affected by the emergence of charge and orbital order without changes in the magnetic ordering. We attribute this change to the lower mobility of $e_g$ electrons in a charge ordered environment.

  \begin{figure}
   \includegraphics[width=0.98\linewidth]{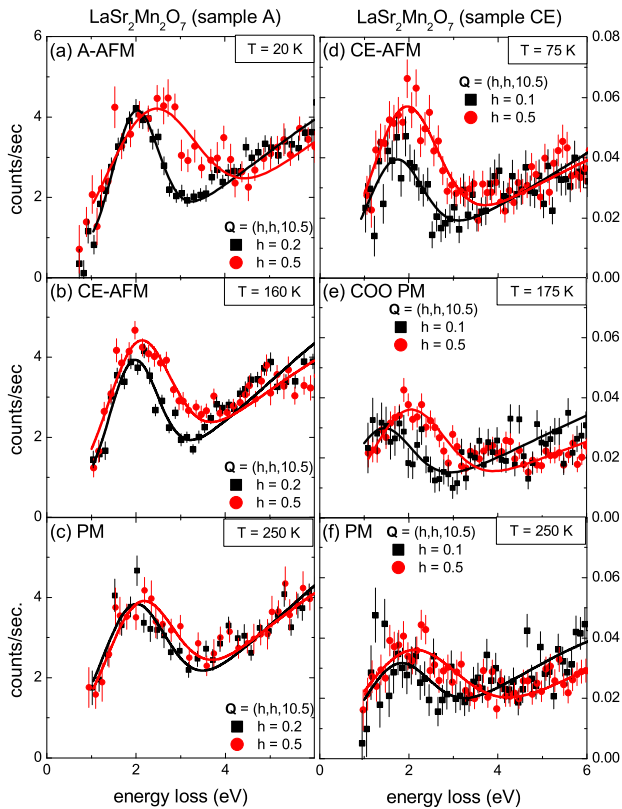}
   \caption{(color online) Background subtracted RIXS spectra in (a)-(c) reentrant \LSMOH  (sample A) and (d)-(f) non-reentrant \LSMOH (sample CE). Temperatures and respective magnetic phases are given in the topline of each panel. For each temperature, one scan taken at $\textbf{Q} = (h, h, 10.5)$, $h = 0.2$ (sample A) or $0.1$ (sample CE), (squares) and one taken at $h = 0.5$ (dots) are shown. Solid lines are the results of the three-peak fits (see text and figure~\ref{fig_1}).}
   \label{fig_5}
  \end{figure}

  \begin{figure}
   \includegraphics[width=0.9\linewidth]{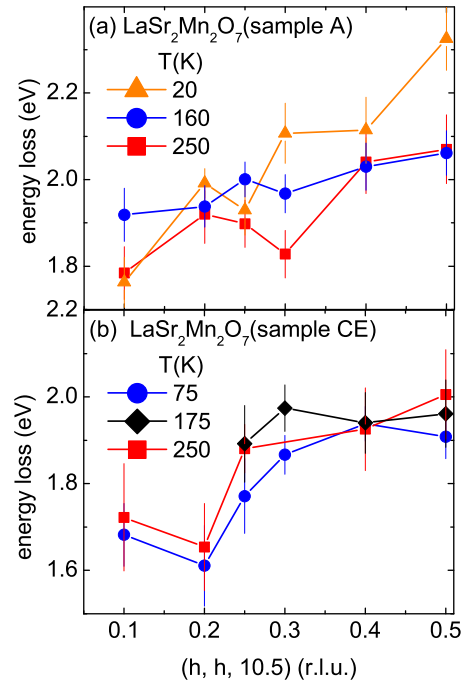}
   \caption{(color online) $\textbf{q}$ and temperature/magnetic phase dependence of the energy loss positions of the low energy excitation in (a) reentrant \LSMOH (sample A) and (b) non-reentrant \LSMOH (sample CE). Lines are guides to the eye. Points for $h = 0.1$ and $0.2$ at \wert{T}{175}{K} are not shown in (b) because a fit with independent parameters was not possible as explained in the text.}
   \label{fig_6}
  \end{figure}

We conclude that our evaluation of the RIXS intensities of the energy loss region \rangeS{1}{5} as function of magnetic order is qualitatively explained by the intersite $d-d$ excitation model\cite{Grenier05}. However, due to the layered structure of \LSMOE, the in-plane magnetic order rather than the three dimensional one has to be taken into account. The observed loss of influence of the magnetic order for small $\textbf{q}$ remains an open question and more measurements, e.g. in other crystallographic directions, are needed.

As a last but very important point, we turn to the $\textbf{q}$ dependence of the peak energy as fitted by the Gaussian lines, e.g. shown in figure~\ref{fig_2}. In order to illustrate the wave vector dependence of the $d-d$ excitation, we show RIXS spectra at small wave vectors, i.e. $h = 0.1$ or $0.2$, overlaid with spectra taken at $h = 0.5$ for every temperature measured in the half-doped samples in figure~\ref{fig_5}. We note that the large difference in count rates between the two samples is due to the usage of a newly manufactured Mn Analyzer\cite{Casa10}  for the measurements on sample A. As the intensity of the $d-d$ excitation increases for decreasing temperatures, we also observe more pronounced wave vector dependencies of the scattering in both samples. In particular, we see a substantial upward dispersion of the peak center of mass of about $0.5\,\rm{eV}$ in the A-type antiferromagnetic phase of sample A. To our knowledge, such a wave vector dependence in RIXS has not been reported before in manganites. It is a direct experimental proof of the non-local character of the observed $d-d$ excitation.

Effects at other temperatures are smaller, except maybe in the charge and orbital ordered paramagnetic phase in sample CE. However, for this particular temperature we could not distinguish the low energy loss side of the excitation peak at $h = 0.1$ and $0.2$ from the elastic line. Thus, the fit shown in figure~\ref{fig_5} was only possible with a fixed linewidth of $2.5\,\rm{eV}$, which is an average value of fitted linewidths at other temperatures and/or wave vectors. Therefore, results for the dispersion at these wave vectors and temperature are not included in the summary presented in figure~\ref{fig_6}. However, energy loss positions are certainly not higher than at other temperatures at the same wave vectors (see figure~\ref{fig_5}).

Figure~\ref{fig_6} shows a more or less pronounced upward dispersion in all phases. However, while the dispersions differ significantly between different phases of sample A, they are identical within the experimental error for all three temperatures measured in sample CE. We interpret these results in the frame work of double exchange. There, the conductivity is mainly characterized by the hopping rate $t$ of the $e_g$ electrons, which is proportional to the angle $\eta$ between two neighboring Mn spins, $t \propto {t_0\, cos(\frac{\eta}{2})}^2$. Thus, the increased dispersion in the A-type antiferromagnetic phase reflects a high in-plane mobility in agreement with a reported high, metallic-like conductivity in this phase\cite{Li07}. Differences between the different phases in sample CE might be much smaller and therefore undetectable, as phases at all three temperatures show insulating behavior.

\section{Conclusions}
We report a detailed investigation of the electronic excitations in various different ordered phases of \LSMOE, $x = 0.36$ and $0.5$, in the energy range of \rangeS{1}{15}. Using resonant inelastic x-ray scattering at the Mn K-edge we probed the detailed wave vector dependence along the crystallographic $(110)$ direction. Our observation of a dispersion as large as $0.5\,\rm{eV}$ within $\textbf{q} = (h, h, 0)$, $h = 0.1 - 0.5$, is a direct, model-independent proof of the non-local character of the corresponding dispersion. Overall, we see qualitative agreement between observed intensity variations of this excitation in different magnetic phases and the corresponding model prediction based on an intersite $d-d$ excitation, originally proposed for pseudo-cubic manganites\cite{Grenier05}. However, the in-plane rather than the three dimensional magnetic order has to be taken into account due to the two-dimensional nature of \LSMOE. Only at two wave vectors, i.e. $h = 0.1$ and $0.2$, in the reentrant sample A we did not see a detectable intensity differences between ferromagnetic and paramagnetic in-plane ordering. This could be related to the particularly strong phase competition in this sample but more data are needed in order to solve this question.

Work at Argonne National Laboratory was supported by the U.S. Department of Energy, Basic Energy Sciences-Materials Sciences, under Contract No. DE-AC02-06CH11357.

\end{document}